# Robust Machine Learning for Regulatory Sequence Modeling under Biological and Technical Distribution Shifts

**Name: YIYAO YANG**
**Email: yy3555@tc.columbia.edu**
**Affiliation: Columbia University, New York, NY, United States**
**ORCID: 0009-0001-8693-4888**

**Abstract**

The robustness of machine learning for regulatory genomics is studied under biologically and technically induced distribution shifts. Deep convolutional and attention-based architectures show good in-distribution performance for deoxyribonucleic acid (DNA) sequence prediction for regulatory genomics, although they are tested under independent and identically distributed (i.i.d.) conditions, whereas in practice, cell-type-specific programs, evolutionary conservation, assay protocol changes, and sequencing artifacts are encountered. We propose a robustness evaluation methodology using a simulation benchmark with mechanistic models and empirical evaluation using a massively parallel reporter assay (MPRA) dataset to assess performance degradations, calibration, and reliability using uncertainty measures. We simulate regulatory outputs using motif-based programs with cell-type specificity, position weight matrix (PWM) perturbations, guanine–cytosine (GC) content bias, sequencing depth, batch effects, and heteroscedastic noise, and evaluate convolutional neural network (CNN), bidirectional long short-term memory (BiLSTM), and transformer architectures under distribution shifts. The models perform well and remain well-calibrated under GC content shifts, although error rates increase and variance miscalibration and coverage collapse are observed under motif-effect rewiring and noise, indicating robustness issues not observed under i.i.d. We also show the effect of incorporating biological structural priors, such as motif-based features and GC content, for improved out-of-distribution (OOD) performance and robustness, although improvements are limited under strong assay noise settings, indicating robustness improvements under genomic shifts. Selective prediction using uncertainty is also studied as a safety net, and risk coverage is evaluated, indicating improved reliability under shift settings, including GC-based OOD settings, although reliability improvements are reduced under noise settings.

## II. Highlight

(1) We propose a machine learning framework for robustness under distribution shifts.

(2) We demonstrate that motif-informed priors improve OOD generalization.

(3) We identify reliability loss under biological and technical shifts.

(4) We establish selective prediction as a safety mechanism under OOD.

(5) We evaluate robustness and uncertainty using simulation and real MPRA data.

## III. Introduction & Literature Review

Understanding the relationship between DNA sequences and their regulatory functions is an essential problem in computational genomics, with applications in enhancer regions, variant effect prediction, and disease genetics [1]. Recent advances in deep learning, specifically convolutional networks and models with long-range genomic context using attention mechanisms, have significantly improved the accuracy of models in predicting regulatory functions, including chromatin accessibility, TF binding, and gene expression prediction using raw DNA sequences [2, 3, 4]. Although the use of these models has become more prevalent in scientific and clinical applications, their evaluation remains mostly limited to in-distribution evaluations, with assumptions of independence and identical distribution. However, the reality of regulatory genomics is that the distribution of data is far from being independent, given the natural biological and experimental variations that are non-i.i.d. with respect to the original distribution of the data.

Although the use of these models in regulatory genomics is becoming more prevalent, their robustness properties are yet to be explored. The early use of deep learning models in regulatory genomics was primarily based on the use of convolutional neural networks to predict TF binding, chromatin regions, and other regulatory functions using raw DNA sequences [2, 5, 6]. The use of convolutional networks was later improved with the extension of the genomic context using longer DNA sequences, with some models using attention mechanisms to improve the accuracy of variant effect prediction using DNA sequences [7, 4].

Apart from accuracy on epigenomic tracks, other applications of sequence models include inference of enhancer-promoter interaction [8, 1], saturation mutagenesis, and massively parallel reporter assays for estimating mutation effect [9], as well as variant prioritization in association studies and fine-mapping experiments [10, 11]. This shows an emerging trend of using sequence models as an alternative to experimental approaches, with potential applications in experimental design and variant interpretation.

While significant progress has been achieved in the evaluation of existing models, existing evaluation protocols are mostly limited to i.i.d. settings, often using random genomic splits or chromosomes for in-distribution generalization. However, in actual regulatory genomics problems, distribution shifts are significant, including cell-type-specific regulatory programs [12], species divergence, and evolutionary turnover of enhancers [13], as well as differences in experimental protocols, readouts, and technical artifacts arising from sequencing depth and batch effects. In machine learning parlance, these are examples of domain shift, covariate shift, and structured OOD problems, which are well-studied in computer vision and natural language processing [14].

Similarly, whereas uncertainty quantification and calibration are now an essential part of ML in high-stakes decision-making, via Bayesian approximations [15], ensembling [16], and selective prediction, these techniques are not well explored in genomic sequence models. Regulatory modeling studies rarely investigate calibration, robustness under OOD conditions, or performance of selective prediction, despite the increasing practice of using model outputs in decision-making. Therefore, collectively, existing sequence-based regulatory models have high predictive accuracy but no systematic assessment or mitigation of robustness, uncertainty, or reliability in the face of biological or technical distribution shifts. This calls for a robust framework that is specifically geared towards regulatory genomics.

## IV. Research Questions

**Research Question 1:** To what extent do current regulatory sequence models preserve predictive performance and calibration under biologically and technically induced distribution shifts?

**Research Question 2:** Can incorporating biological structural priors improve robustness under distribution shift without degrading in-distribution performance?

**Research Question 3:** Can uncertainty estimation and selective prediction provide biologically meaningful safety mechanisms when regulatory sequence models encounter out-of-distribution (OOD) genomic conditions?

## V. Research Techniques & Programming Tools

The programming language used is Python. For every research question, the different methods of computation and Python libraries used across the various research questions (Table 1), from sequence-to-function convolutional architectures and PWM-based structural priors to uncertainty-aware selective prediction, are mentioned along with the Python libraries used for the implementation of the methods. This makes the computation transparent and reproducible for the analysis performed.

**Table 1**
*Computational Methods and Python Libraries Across Research Questions*

| Research Questions | Technique Used | Python Libraries |
|---|---|---|
| Peroformance & Calibration under Shifts | - Sequence-to-function modeling (CNN)<br>- Biological covariate shifts (GC)<br>- Technical shifts (noise)<br>- Regression metrics (MSE)<br>- Calibration evaluation (ECE) | - PyTorch (CNN regression & classification)<br>- Scikit-learn (performance metrics)<br>- Numpy (GC-biased sequence simulation & array operations)<br>- SciPy (Gaussian quantiles for coverage)<br>- Matplotlib & Seaborn (performance and calibration plots) |
| Structural Priors for Robustness | - CNN + PWM structural priors<br>- Sequence–motif feature fusion<br>- Heteroscedastic Gaussian regression<br>- Robustness evaluation across genomic shifts<br>- Variance calibration (Var-ECE, coverage) | - PyTorch (model architecture)<br>- Numpy (data processing)<br>- Scikit-learn (evaluation metrics)<br>- Matplotlib & Seaborn (visualization) |
| Uncertainty & Selective Prediction | - Heteroscedastic variance–based uncertainty<br>- Confidence-based selective prediction (risk–coverage)<br>- Selective risk analysis (regression and classification)<br>- Robustness under biological and technical OOD shifts<br>- Selective abstention across shifts | - PyTorch (CNN baseline model and batch inference)<br>- NumPy (uncertainty-based coverage filtering and sample ranking)<br>- Scikit-learn (classification accuracy and risk estimation)<br>- Matplotlib and Seaborn (risk–coverage curves and heatmap visualizations for regression and classification) |

*Note.* Programming Language: Python

## VI. Experimental Setup

### (VI.1) Experimental Setup: Synthetic Sequence Simulation

#### (VI.1.1) Benchmark Environments and Dataset Generation

We synthesized DNA sequences of length $L = 1000$ using $K = 5$ randomly generated Position Weight Matrices (PWMs). Five distinct environments were constructed to evaluate model robustness:

- **In-distribution Baseline:** DNA sequences generated with GC-bias $g = 0.6$, baseline motif scalar alpha base, and homoscedastic technical noise ($\sigma = 0.3$).
- **Biological GC-Shift:** Sequences generated with an altered GC-bias $g = 0.3$, introducing compositional distribution shift.
- **Technical Heteroscedastic Noise Shift:** Baseline GC-content ($g = 0.6$) subjected to strong heteroscedastic signal-dependent noise ($\sigma = 0.8$), where noise intensity scales with motif activation magnitudes.
- **Biological Concept Shift:** Biological rewiring of motif expression weights, shifting the underlying mapping between motifs and continuous targets.
- **Combined Shift:** Simultaneous application of GC-bias shift ($g = 0.3$), concept rewiring, and heteroscedastic noise ($\sigma = 0.8$).

**(VI.1.2) Model Architectures and Training Protocols**

The Heteroscedastic Regression Model is built upon a 1D Convolutional Neural Network (1D CNN) backbone to capture high-dimensional features from time-series or one-dimensional signals. The model accepts an input with 4 channels, which undergoes feature extraction through a 1D convolutional layer with a kernel size of 15, padding of 7, and an output channel size of 32. Next, an Adaptive Average Pooling layer reduces the spatial dimension to a fixed size of 64. The extracted features are then fed into a dense bottleneck layer that compresses the 3200-dimensional flattened representation ($32 \times 64$) down to 64 dimensions using a ReLU activation function. At the network's head, dual output branches operate in parallel to predict the Gaussian distribution mean $\hat{\mu}$ and log-variance $\log \hat{\sigma}^2$, respectively. The log-variance predictions are clamped to $[-5.0, 5.0]$ for numerical stability. Prior to training, the target labels were standardized to zero-mean and unit-variance. The model is trained for 15 epochs using the Adam optimizer with a learning rate of $10^{-3}$ and batch size of 64, minimizing a Gaussian Negative Log-Likelihood (NLL) loss.

**(VI.1.3) Selective Prediction Confidence Proxies**

In heteroscedastic regression where the target variable $y$ follows a Gaussian distribution $\mathcal{N}(\mu(x), \sigma^2(x))$, sample confidence is defined as the negative predicted variance: $C_{regression}(x) = -\sigma^2(x) = -\exp(s(x))$. Here, $C_{regression}(x)$ is an inverted measure of aleatoric uncertainty used to rank low-variance predictions first when constructing Risk-Coverage MSE curves across varying coverage thresholds. Predictions with lower variance yield higher confidence scores. During evaluation, samples are sorted in descending order of $C_{regression}(x)$, prioritizing low-uncertainty samples with the lowest variance, to construct Risk-Coverage Mean Squared Error (MSE) curves, thereby systematically evaluating model performance under varying data coverage thresholds.

For binary classification tasks with predicted output probability $p = P(y = 1 | x)$, instance confidence is defined as the maximum likelihood over predicted classes: $C_{classification}(x) = \max(p, 1 - p)$. This score measures the model's certainty regarding its decision, where values closer to 0.5 denote high uncertainty and values closer to 1 indicate strong conviction. Model risk is evaluated by calculating the classification error rate across varying coverage thresholds, selecting only samples whose confidence meets or exceeds each threshold.

**(VI.2) Experimental Setup: Real-World Genomic Data Benchmark**

**(VI.2.1) Dataset Specifications and Distributional Partitioning**

Experiments were conducted on high-throughput genomic HDF5 benchmark datasets (train.hdf5, valid.hdf5, test.hdf5) in the Massively Parallel Reporter Assay (MPRA) repository from Stanford University, via the public index at mitra.stanford.edu/kundaje/projects/mpra/data [17]. Inputs consist of 145 bp one-hot encoded DNA sequences (shape: 145 x 4) predicting 12 continuous expression/binding targets (shape: $N$ x 12). To benchmark model calibration under real-world shifts, validation sequences were partitioned into:

- **In-distribution Reference:** The overall validation set consisting of 19,833 sequences.

- **Biological GC-Shifts:** Sequences partitioned by GC-content ratio into Mid-GC (35% to 65%, 4,363 samples), Low-GC ( $< 35\,\%$ , 163 samples), and High-GC ( $> 65\,\%$ , 474 samples).

- Technical Shifts: Gaussian noise added directly to one-hot inputs, clipped to [0,1] at intensity levels $\sigma = 0.2$ and $\sigma = 0.4$ to simulate sequencing measurement noise.

**(VI.2.2) Model Architectures and Calibration Metrics**

The baseline model, designated as SimpleSeqCNN, employs a sequence-only convolutional architecture designed to process input sequences through three consecutive one-dimensional convolutional layers. These layers progressively expand the feature channels from 64 to 128 and finally to 128, utilizing a uniform kernel size of 7 and padding of 3 to preserve spatial dimensionality. Each convolution is coupled with standard Batch Normalization and ReLU activation, followed by a dropout rate of 0.3 to prevent overfitting. Global average pooling is then applied across the sequence length dimension to collapse spatial features, which are subsequently fed into a dense linear head that projects the 128-dimensional pooled representations down to the 12 target outputs.

To incorporate sequence-level compositional biases, the structural-prior variant (GCStructCNN) extends the baseline by utilizing a dual-branch architecture. While the sequence branch mirrors the baseline to extract a 128-dimensional visual feature representation, a dedicated global GC-content multi-layer perceptron (MLP) branch processes scalar GC composition through sequential dense transformations (1 to 32 to 32 dimensions) with ReLU activations. The extracted 128-dimensional sequence features and 32-dimensional GC features are concatenated into a unified 160-dimensional vector prior to the final linear projection layer to produce the 12 target predictions.

Both models are trained under an identical protocol using the Adam optimizer with a learning rate of 0.001, a weight decay of $1 \times 10^{-5}$, and a batch size of 32, optimizing a Mean Squared Error (MSE) loss function. Model performance and output calibration are evaluated using MSE alongside linear regression calibration parameters.

**(VI.2.3) Random Seeds & Reproducibility Protocol**

To ensure full reproducibility of all simulated sequence generation, dataset shuffling, model parameter initialization, and technical noise injection, all pseudo-random number generators across Python, NumPy, and PyTorch were initialized with a fixed global random seed (seed = 42). Quantitative performance evaluation metrics across all benchmark environments correspond to standardized, deterministic runs under this controlled seed setup.

## VI. Simulated Regulatory Benchmark

In order to investigate robust properties of models of regulatory sequences under controlled shifts in distributions, we suggest a simulated benchmark where we take into consideration some of the biological and technical factors involved in gene regulation. This is unlike real-world genomics datasets, where biological variability and technical artifacts are confounded. Moreover, unlike real-world datasets, where uncertainties underlying ground truth are inaccessible, our simulation framework offers us explicit control over cell-type-specific regulatory logic, technical variability, and accessible heteroscedastic noise.

### Regulatory Sequence Generation

We simulate a DNA sequence $x \in \{\mathrm{A, C, G, T}\}^{L}$ of fixed length $L = 1000$. Background nucleotides are sampled from a GC-biased distribution reflecting variable chromatin composition:

$P(A) = P(T) = \frac{1-g}{2}, P(C) = \frac{g}{2}$, where $g \in [0,1]$ contains GC-content and can later induce GC-related technical shifts.

### Motif-Based Regulatory Activity

We define $K$ as a transcription factor (TF) motif $\{M_k\}_{k=1}^{K}$, simulated as position weight matrices (PWMs).

For each sequence, motif occurrences contribute to additive activation:

$$a_k(x) = \sum_{t=1}^{L-|M_k|+1} \mathbf{1}\left[x_{t:t+|M_k|-1} = M_k\right] \cdot w_k, \text{ where } w_k > 0 \text{ represents motif affinity strength.}$$

This modeling choice reflects regulatory logic observed in enhancer sequences, in which the presence and multiplicity of TF binding motifs contribute to regulatory output.

**Cell-Type-Specific Regulatory Logic**

To simulate biologically meaningful variation, we define $C$ cell types with TF activity coefficients $\alpha_{c,k}$.

Regulatory output for cell type $c$ is $z_c(x) = \sum_{k=1}^{K} c_k \, a_k(x)$.

We consider two readout regimes:

- Continuous: $y_c(x) = z_c(x) + \epsilon_c, \qquad \epsilon_c \sim \mathcal{N}\left(0, \sigma_c^2(x)\right)$
- Binary: $y_c(x) = \mathbb{I}\left[z_c(x) > \tau_c\right]$

Since $\sigma_c(x)$ varies by motif content, the continuous regime enables heteroscedastic uncertainty, which is essential for calibration experiments.

**Biological Distribution Shifts**

We simulate biological OOD scenarios by varying TF activity coefficients across train/test cell types. Let the training cell types be $C_{train}$, and OOD test cell types be $C_{test}$, with $C_{\text{train}} \cap C_{\text{test}} = \emptyset$.

We sample TF activities as: $\alpha_{c,k}^{\text{test}} \sim \mathcal{N}\left(\alpha_{c,k}^{\text{train}}, \sigma_{\text{bio}}^2\right)$, where $\sigma_{bio}$ controls the magnitude of biological domain shifts. Larger $\sigma_{bio}$ corresponds to stronger tissue-specific regulatory divergence analogous to GTEx multi-tissue expression variation.

We further simulate motif evolution via PWM perturbations: $M_k^{\text{test}} = T\left(M_k^{\text{train}}, \delta\right)$, where substitution probability $\delta$ mimics enhancer turnover across species.

**Technical Distribution Shifts**

We simulate biological OOD scenarios by varying TF activity coefficients across train/test cell types. Let the training cell types be $C_{train}$, and OOD test cell types be $C_{test}$, with $C_{\text{train}} \cap C_{\text{test}} = \emptyset$. We sample TF activities as: $\alpha_{c,k}^{\text{test}} \sim \mathcal{N}\left(\alpha_{c,k}^{\text{train}}, \sigma_{\text{bio}}^2\right)$, where $\sigma_{bio}$ controls the magnitude of biological domain shifts. Larger $\sigma_{bio}$ corresponds to stronger tissue-specific regulatory divergence analogous to GTEx multi-tissue expression variation.

We further simulate motif evaluation via PWM perturbations: $M_k^{\text{test}} = T\left(M_k^{\text{train}}, \delta\right)$, where substitution probability $\delta$ mimics enhancer turnover across species.

**Technical Distribution Shifts**

Technical variation is simulated independently of biological logic:

(1) **Sequencing Depth Shift:** We apply multiplicative noise: $\tilde{y}_c = y_c \cdot d, d \sim \text{LogNormal}\left(\mu_d, \sigma_d^2\right)$, modeling read-depth variation in ATAC-seq/RNA-seq.

**(2) Batch Effects:** We add nuisance offsets: $\tilde{y}_c = y_c + b, b \sim \mathcal{N}\left(0,\sigma_b^2\right)$ , capturing between-batch variability in high-throughput assays.

**(3) GC-Bias:** We distort output based on GC-content $\tilde{y}_c = y_c \cdot \left(1 + \beta\left(GC(x) - \mu_{GC}\right)\right.$, reflecting GC-dependent mapping efficiency commonly observed in sequencing pipelines. Together, these perturbations simulate covariate shift, label shift, and spurious correlations frequently seen in functional genomics.

**Structured Perturbations:** For assessing post hoc robustness, we have proposed perturbations, including motif knockout (removal of motif instances $\Delta M = 0$), motif insertion (addition of strong sites), rewiring of TFs (permutes $\alpha_{c,k}$ among all TFs), and motif masking (simulates epigenetic silencing). These operations mimic enhancer deletion, gain-of-function mutations, or rewiring of chromatin structure, all of which are relevant to saturation mutagenesis or MPRA screens.

**Model Architectures:** We test three architectures of interest: the CNN-based regulatory model (inspired by local motif detection), the BiLSTM-based model (inspired by mid-range motif dependencies), and the Transformer-based long-range model (inspired by the Enformer architecture that supports long-range dependencies). The Transformer architecture utilizes sinusoidal positional encodings and multi-head attention; however, the sequence length is reduced for the simulation scale.

**Evaluation Metrics:** We evaluate models along three axes aligned with our research questions:

For RQ1 (Performance under Distribution Shift), the evaluation metrics are In-Distribution (ID) accuracy, Out-Of-Distribution (OOD) accuracy, Worst Group Accuracy, and Domain Gap $\delta = ID - OOD$. The evaluation metrics for RQ2 (Structural Robustness) are Motif Perturbation Sensitivity and TF-Rewiring Stability. The evaluation metrics for RQ3 (Uncertainty & Selective Prediction) are Predictive Entropy, MC Dropout Variance, and Deep Ensemble Variance. Calibration metrics are Expected Calibration Error (ECE), Negative Log-Likelihood (NLL), and Brier Score. The evaluation metrics for Selective Prediction are given by Risk Coverage Curves $R(c) = [l(f(x), y) \,|\, u(x) \geq u_c]$ .

Simulation Summary: The above simulation benchmark, when taken as a whole, represents a controlled but biologically inspired environment for assessing robustness. By explicitly modeling biological variability (in terms of cell-type-specific regulatory programs and motif evolution), as well as various types of technical variability (such as depth shift effects, batch effects, and GC bias), and heteroscedastic ground truth uncertainty, it represents well-defined and highly controlled OOD scenarios that are inspired by biologically motivated but highly specific types of motif-based regulatory logic.

## VII. Data Analysis & Results

### (VII.1) Performance on Simulated Dataset
### Research Question 1: Evaluating Robustness Under Genomic Distribution Shifts

To measure the robustness of our methods, we tested the heteroscedastic Gaussian CNN and its derived binary classifiers under various biologically and technically inspired distributional changes. The changes tested were GC-content covariate shift, heteroscedastic assay noise, motif-effect rewiring (concept shift), and their combination (Figure 1). On the in-distribution test data (Figure 2), our regression model had low error (MSE ≈ 0.103) and well-calibrated variance estimates (Var-ECE ≈ 0.03, 90% coverage ≈ 0.85). Under mild GC-content shift conditions, the model maintained consistent accuracy with a slight increase in error (MSE ≈ 0.107, Var-ECE ≈ 0.04).

**Fig. 1**
*Regression MSE Under Distribution Shifts*

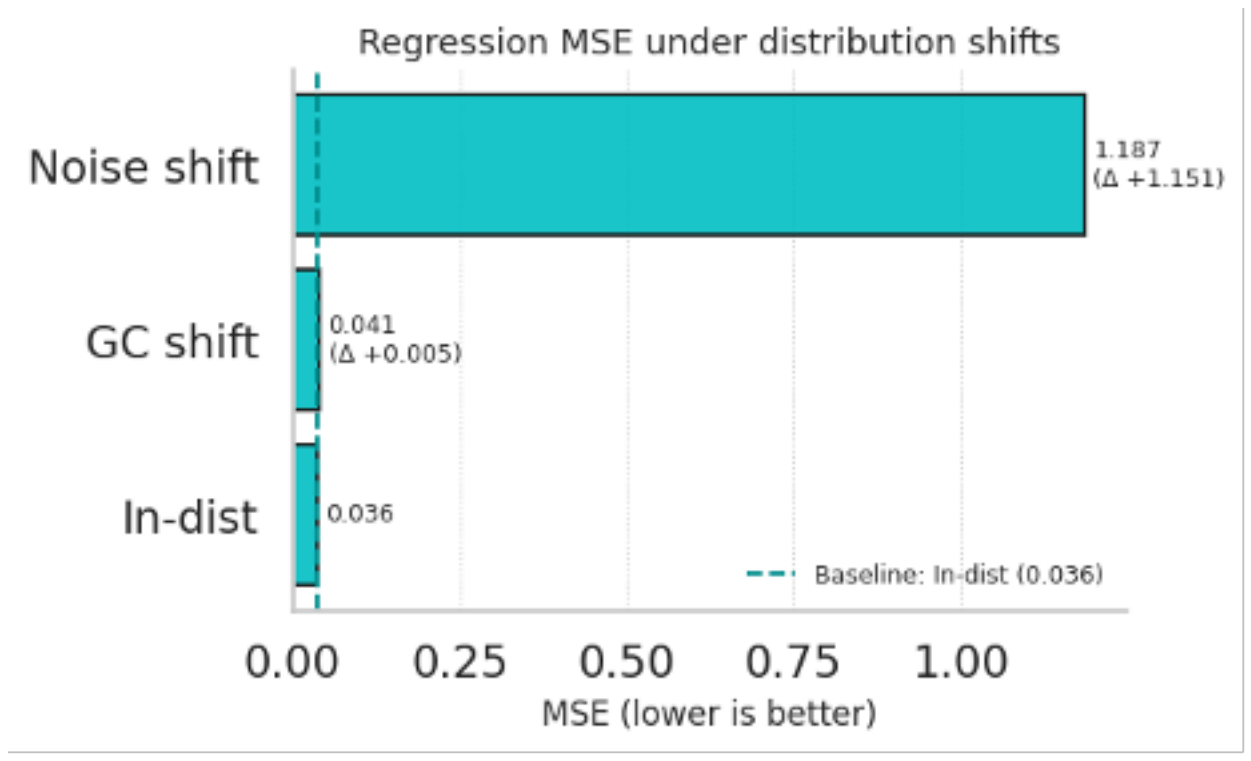

**Fig. 2**
*Variance Calibration Across Environments*

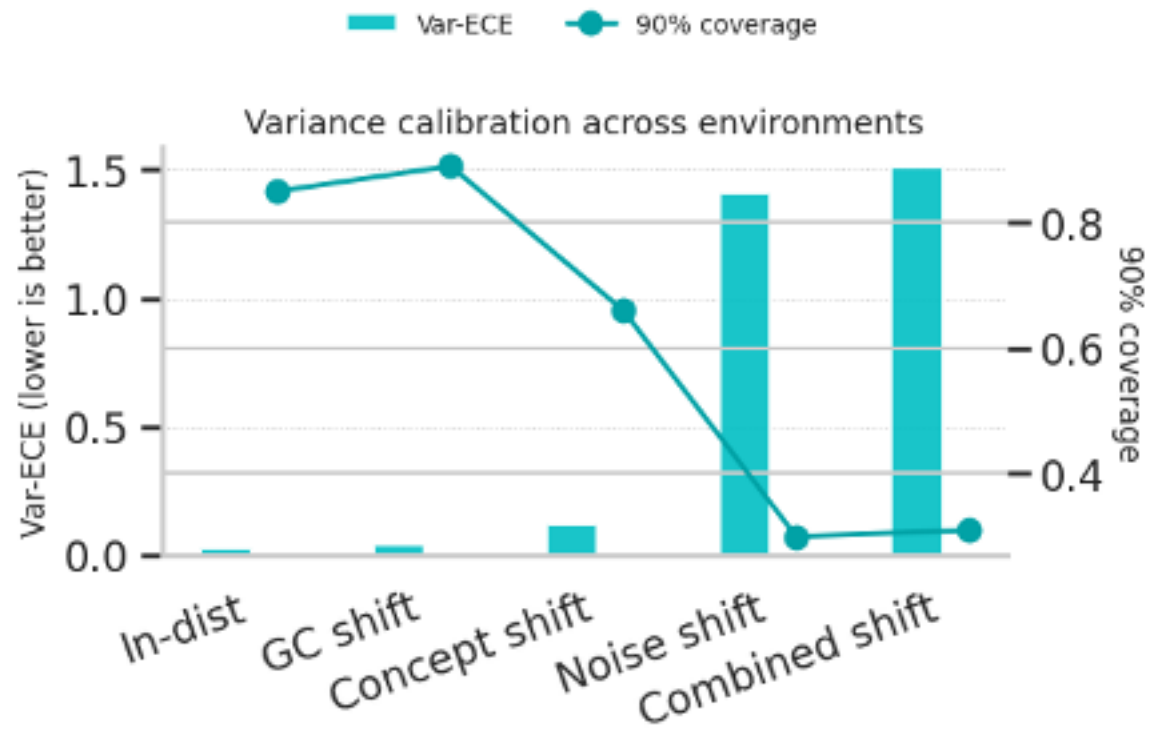

On in-distribution test data, predicted mean activity closely follows true response with data points clustered along the $y = x$ diagonal (Pearson $r = 0.96$) (Figure 3), indicating that the Gaussian regression model obtains a strong baseline before any distribution shift (Figure 4).

**Fig. 3**
*Response on the In-Distribution Test Set*

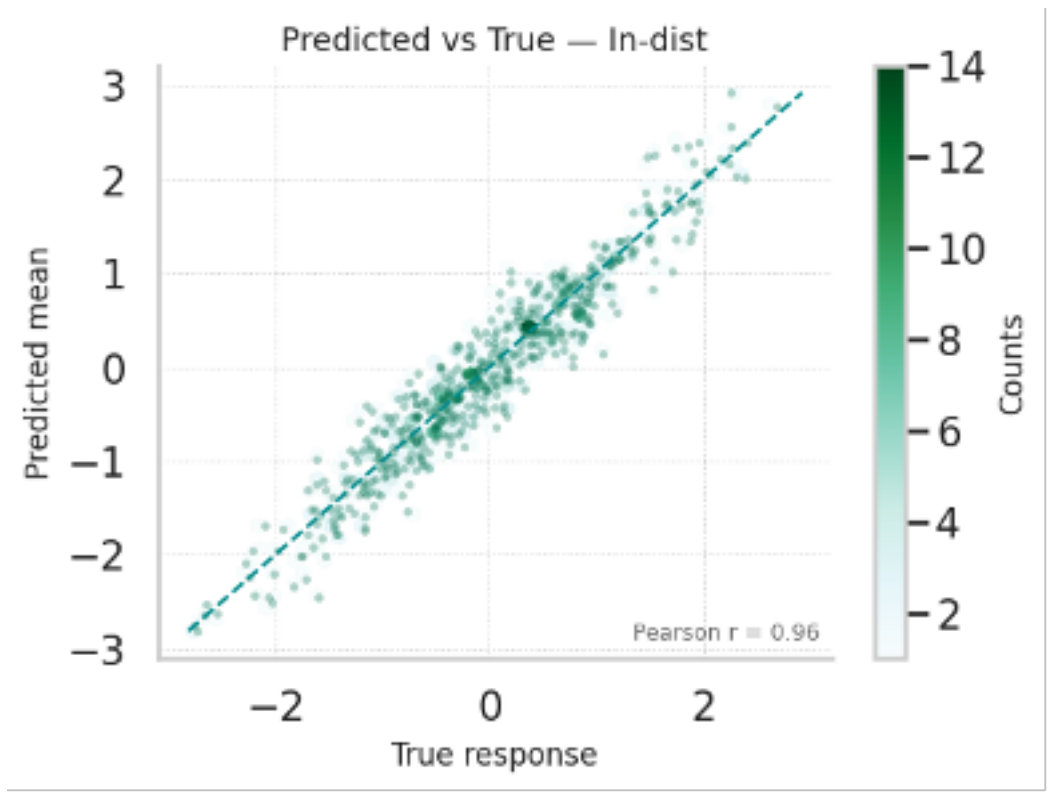

**Fig. 4**
*Residual Distributions*

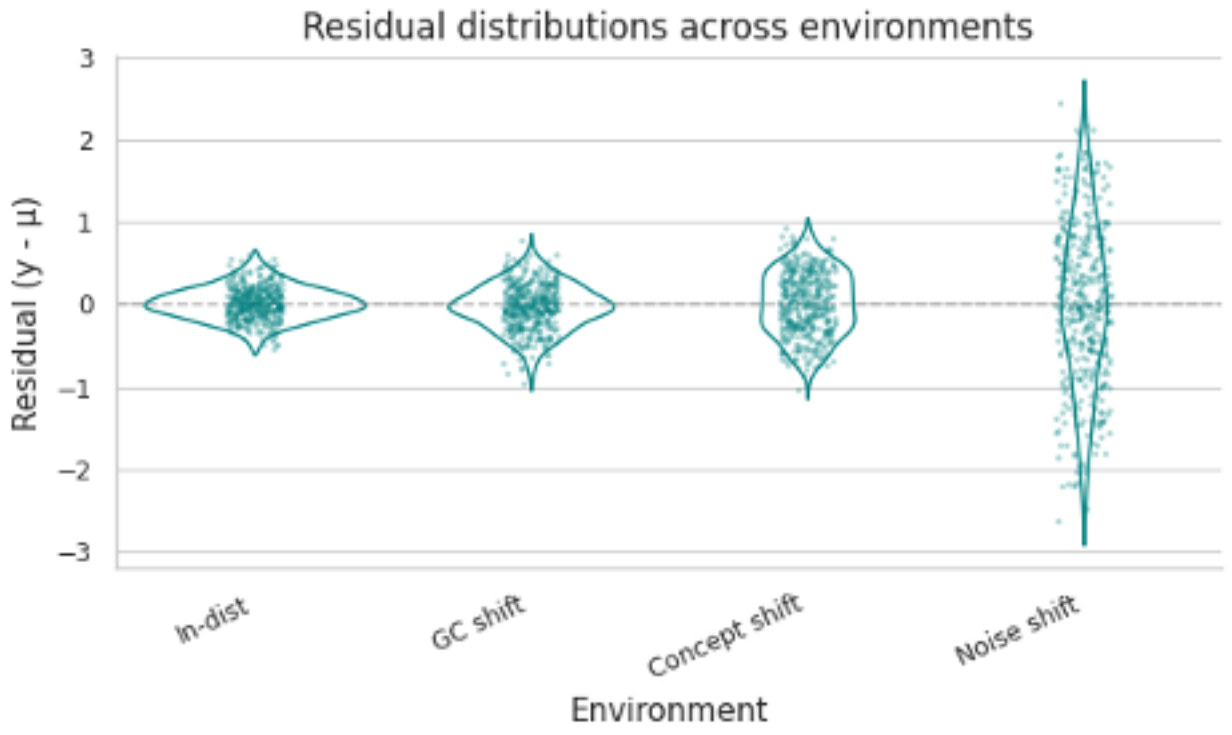

On the other hand, concept shift and technical noise caused significant deterioration, with motif-effect rewiring increasing error rates 2x (MSE ≈ 0.21) and reducing coverage to 0.66, while heteroscedastic assay noise increased error rates over 10x (MSE ≈ 1.50) with significant underestimation of uncertainty (Var-ECE ≈ 1.41, Coverage ≈ 0.30). The joint effect of both shifts caused the most dramatic deterioration in all metrics, with MSE ≈ 1.63, Var-ECE ≈ 1.51. The binary classifier learned from the same embeddings demonstrated analogous behavior, with accuracy varying near chance levels (0.47-0.58) and ECE increasing significantly under concept shift (ECE ≈ 0.34). Overall, the experiments demonstrate that current regulatory sequence models are robust to shallow distribution shifts but become overconfident and error-prone under meaningful mechanism shifts and experimental noise, indicating a new dimension of robustness not captured by IID evaluation.

From a biological point of view, GC-content changes are driven by the compositional differences of genomic regions and species, affecting both nucleosome positioning and TF accessibility, as well as experimental biases in PCR- and sequencing-based assays. Motif-effect rewiring captures the context-dependent behavior of TF binding, with different co-factors, chromatin states, and cell-type programs affecting the regulatory logic without changing the DNA sequence itself. Heteroscedastic noise models the experimental uncertainties in regulatory genomic assays, such as in ChIP-seq, DNase-seq, and ATAC-seq, where coverage, background, and biochemical efficiencies result in noise patterns that depend on the input distribution.

**Research Question 2: Structural-Prior Gaussian Regression**

To effectively incorporate biologically relevant inductive bias into the regression model, we propose using sequence-based CNN representations along with motif-derived structural priors. For an input sequence $x$, we first derive a vector of motif activation scores $m = f_{PWM}(x) \in R^K$, where each $m$ represents the overall match strength of a motif $k$. In parallel, the input sequence $x$ is encoded using a one-dimensional convolutional neural network encoder to derive a representation vector $h_{seq}(x) = \phi_{CNN}(x) \in R^{d_{seq}}$.

The motif scores are passed through a small MLP to obtain a structural embedding $h_{motif}(m) = \psi(m) = \sigma(W_m m + b_m) \in R^{d_{motif}}$. The two components are then fused via concatenation: $h(x, m) = [h_{seq}(x) \,||\, h_{motif}(m)] \in R^{d_{seq}+d_{motif}}$, allowing the model to make use of both generic sequence patterns and the biological basis of motif structures. On top of the fused representation, we add a heteroscedastic Gaussian head that predicts the conditional mean and variance: $\mu(x, m) = W_\mu h(x, m) + b_\sigma$, yielding the predictive distribution $y \,|\, x \;\; N(\mu(x, m), \sigma^2(x, m))$. The model is trained end-to-end by minimizing the negative log-likelihood $L_{NLL} = \frac{1}{2}(\frac{(y - \mu(x, m))^2}{\sigma^2(x, m)} + log\sigma^2(x, m) + log(2\pi))$, which encourages both accurate prediction and good uncertainty. This approach directly encodes PWM-based biological priors into the predictive pipeline without limiting the flexibility of the CNN encoder. The model can now benefit from motif-level regulatory structure without limiting its ability to learn further sequence-based information. The addition of motif-based structural priors provides a better in-distribution baseline. The structural prior model also provides better robustness to meaningful GC shifts and concept shifts. The model's MSE is always lower than that of the unconstrained Gaussian CNN. The structural prior model also has better variance calibration. The model's coverage also tends to be closer to the nominal 90% than that of the unconstrained Gaussian CNN. However, under severe heteroscedastic noise, the improvement of the structural prior model over the unconstrained Gaussian CNN is limited. The structural prior model also suffers from the same failure modes as the baseline. The MSE and variance ECE are high. The coverage of the structural prior model remains much lower than nominal.

In all cases, the structural prior Gaussian CNN performs at least as well as, if not better than, the baseline with regard to regression error (Figure 5). On the in-distribution test data, the inclusion of motif-based priors has a small effect, reducing MSE from 0.103 to 0.093, which supports that using biological structure does not hurt IID performance. On the out-of-distribution test data, the structural prior model shows larger improvements under biological shifts. It reduces error under the GC-content shift, from 0.107 to 0.087, and under the motif-effect rewiring shift, from 0.209 to 0.184, suggesting that this model is more robust under shifts in the underlying biological regulation. However, under severe heteroscedastic assay noise, as well as the combined shift, large errors for both models are observed, with structural priors only providing small improvements from 1.498 to 1.481, as well as from 1.628 to 1.549, suggesting that architectural priors cannot solve issues with technical noise.

In terms of calibration, structural priors provide clear benefits under biological regime shifts. However, structural priors fail to address failures under technical noise (Figure 6). Under in-distribution data, structural priors provide better calibration by achieving lower variance calibration error (Var-ECE: 0.028 to 0.007) along with better coverage (0.85 to 0.90), which suggests that structural information helps the model better quantify uncertainty without degrading coverage. These improvements are also observed under GC content shift with low Var-ECE (0.040 to 0.007), where empirical 90% coverage remains tightly aligned with nominal. Under concept shift, structural priors again provide better calibration by achieving lower miscalibration (0.122 to 0.094) along with better coverage (0.67 to 0.72), which suggests that biological feature fusion helps the model better quantify uncertainty under concept shift. However, under heteroscedastic assay noise and combined shifts, both models are severely miscalibrated (Var-ECE $\geq$ 1.39) with under-coverage ($\approx$ 0.30), which suggests that structural priors are not sufficient to address uncertainty failures under technical noise.

**Fig. 5**
*Regression MSE: Baseline vs Structural-Prior*

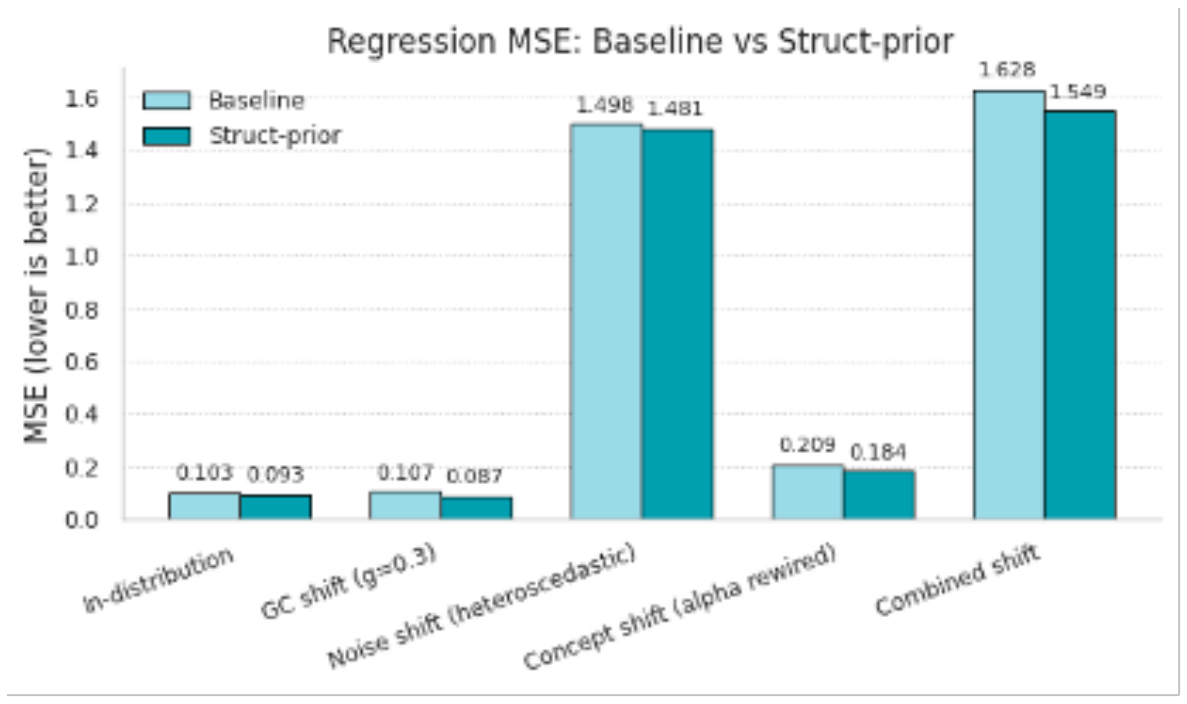

**Fig. 6**
*Variance Calibration & Coverage Under Distribution Shifts*

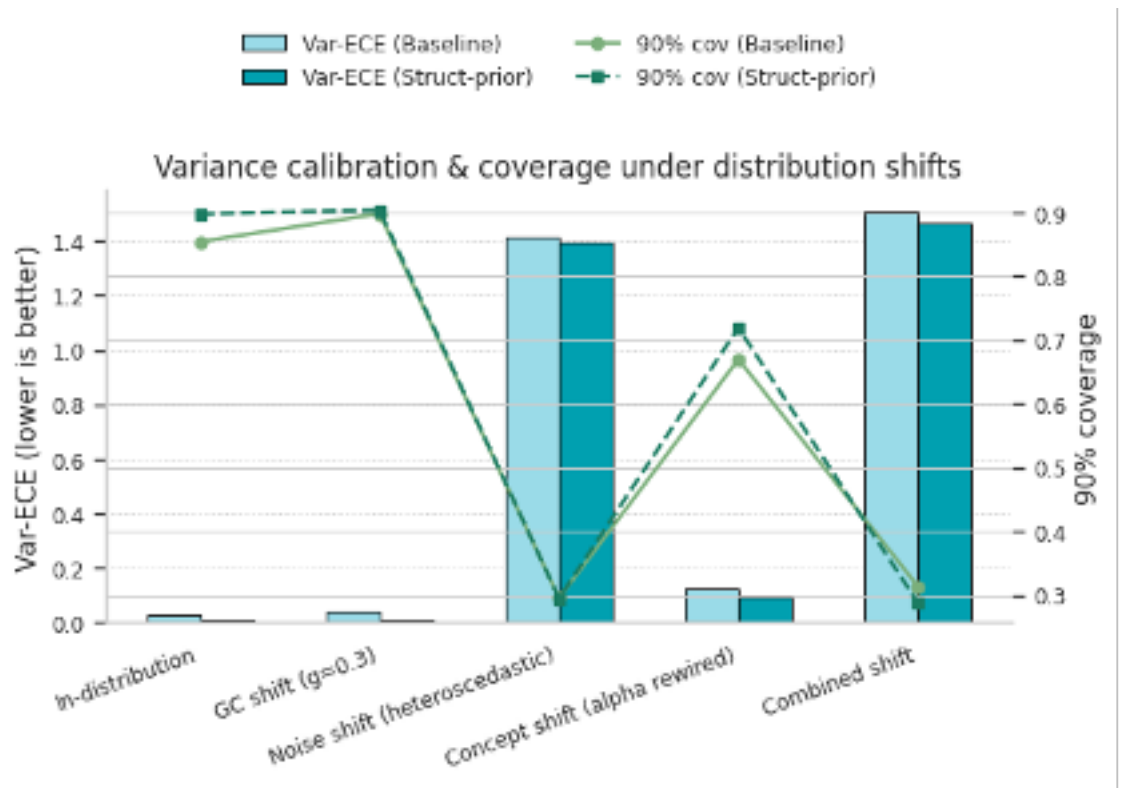

**Research Question 3: Uncertainty-Aware Selective Prediction as an OOD Safety Mechanism**

To test whether such an approach can leverage uncertainty estimates as a means of providing selective safeguards against OOD conditions, we also explored the behavior of the risk-coverage curves for all types of distribution shifts (Figure 7). Under test conditions that are within the distribution and a mild form of GC-content covariate shift, the model exhibits a clear monotonic trend: restricting the model's predictions to the top 10-30% of the most confident predictions reduces the error (MSE ≈ 0.10-0.06) substantially, indicating that the model's variance does indeed correlate well with its estimation accuracy in these conditions. Under the biological mechanism shift of motif-effect rewiring, the approach of selective prediction also partially succeeds, where the low-coverage regime achieves a level of risk close to the in-distribution regime (MSE ≈ 0.18), despite the high level of error at full coverage. In the case of heteroscedastic assay noise and the combination of all shifts, the risk-coverage curves are flat and > 1.3 MSE at low coverage levels, indicating that the model's aleatoric uncertainty has lost its relationship with its correctness in these conditions.

The approach of selective prediction can indeed be an effective means of providing model safeguards against OOD conditions that are biologically relevant. In the case of concept shift ( Figure 8), the performance of both models exhibits the canonical risk-coverage behavior: restricting the model's predictions to the top 10-30% of the most confident predictions reduces the regression error substantially, indicating that the model's variance does indeed correlate well with its correctness even when the regulatory logic of the model has changed. However, the structural prior model achieves a consistently lower level of risk at all levels of coverage (≈ 0.17-0.19 MSE vs. ≈ 0.21-0.22 MSE of the baseline model), indicating that the use of motif-informed representations not only improves the model's robustness against mechanism shifts but also the approach of using uncertainty as a means of model safeguarding.

**Fig. 7**

*Regression Risk-Coverage Under Distribution Shifts*

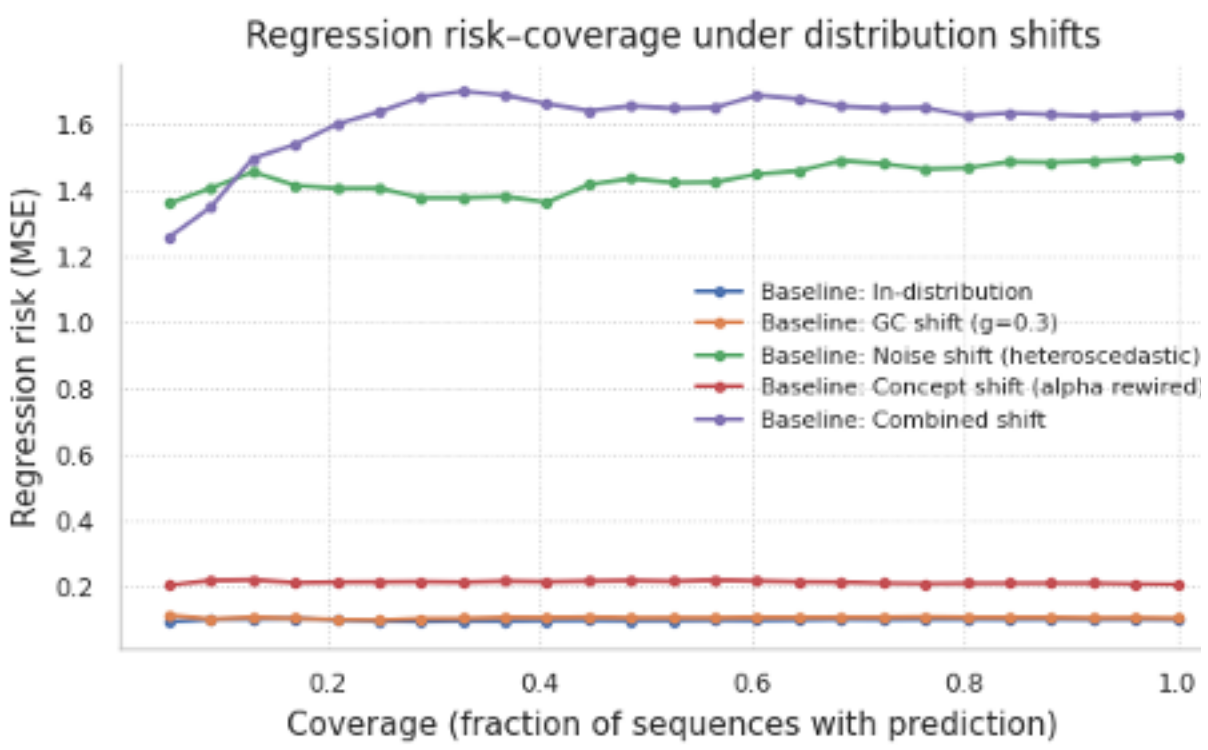

**Fig. 8**

*Risk-Coverage Comparison (Concept Shift)*

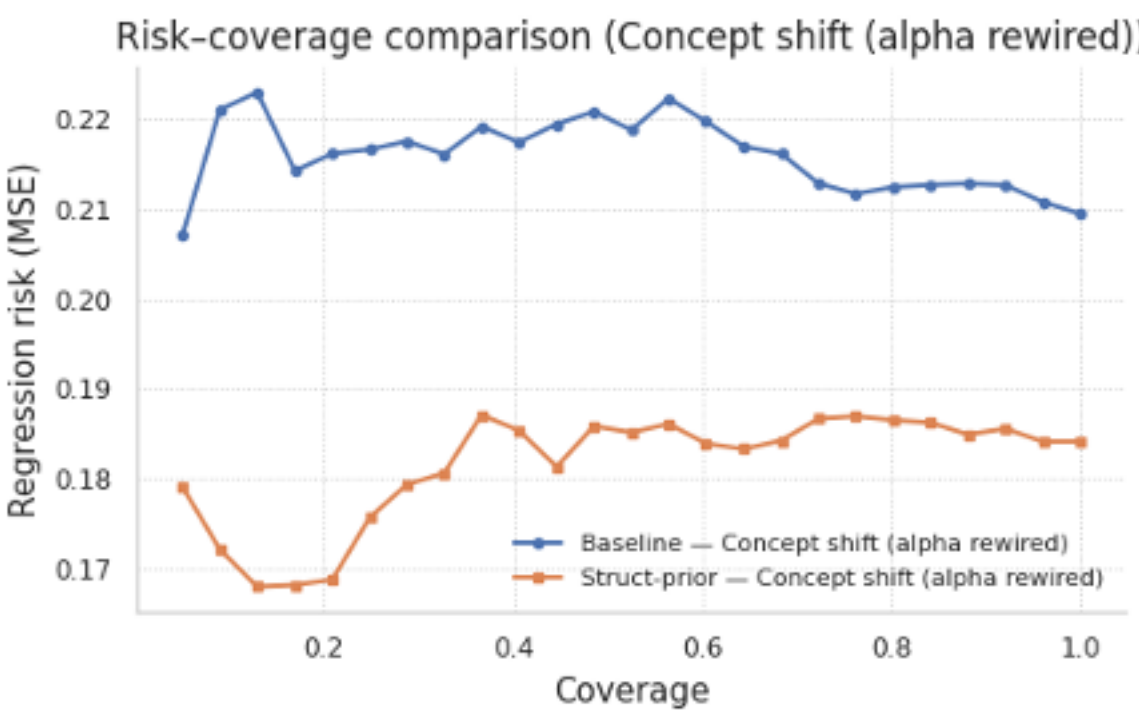

To assess whether uncertainty might facilitate effective abstention in the binary case, we also calculated the risk-coverage curves for classification under the same range of distribution shifts (Figure 9). For all environments, we see that, across the range, restricting predictions to the top 10-30% of predictions by confidence monotonically reduces misclassification risk, suggesting that confidence is still somewhat predictive even under covariate, mechanism, and technical shifts. However, we also see that this effect is highly dependent on the shift type. For instance, under in-distribution and GC-content shifts, we see that some improvement is gained by selective prediction, likely because classification accuracy is still reasonably well-conserved under these shifts. For concept shift, we see that selective abstention is more effective, with risk dropping from 0.42 at full coverage to 0.13 at lower coverage, which suggests that confidence still somewhat tracks correctness even under a fundamentally changed regulatory logic. For heteroscedastic noise and combined shifts, however, we see that selective filtering offers little improvement, with risk still being quite high ( > 0.5) across all levels of coverage, likely because confidence is no longer well-related to correctness under a fundamentally changed signal-to-noise ratio. For binary regulatory decisions, uncertainty is a useful quality control signal under genomic covariate shifts and mechanism shifts, but it fails to detect unreliable decisions under severe technical noise, where assay-level noise overwhelms sequence-level signal. In summary, our results demonstrate that abstaining under uncertainty can be an effective safety mechanism under biologically relevant mechanism shifts, but it fails in OOD scenarios where severe technical noise overwhelms confidence-accuracy alignment.

The selective classification risk (Figure 10) under varying coverage levels for five environments. Higher coverage means fewer abstentions. In the in-distribution environment, as well as under shallow GC-content shifts, risk increases smoothly with coverage, indicating that uncertainty is useful for abstaining from predictions. Under the concept shift, risk is high for low coverage since we can isolate low-risk predictions at low coverage, but risk increases sharply with coverage since we are forced to make predictions on all instances, including those with high uncertainty. Under severe heteroscedastic noise, there is no separation, with risk high across all coverage levels, since noise is high in all instances, making it difficult to abstain from making predictions. In the combined shift, we see behavior similar to that under severe heteroscedastic noise.

**Fig. 9**

*Classification Risk-Coverage Under Distribution Shifts*

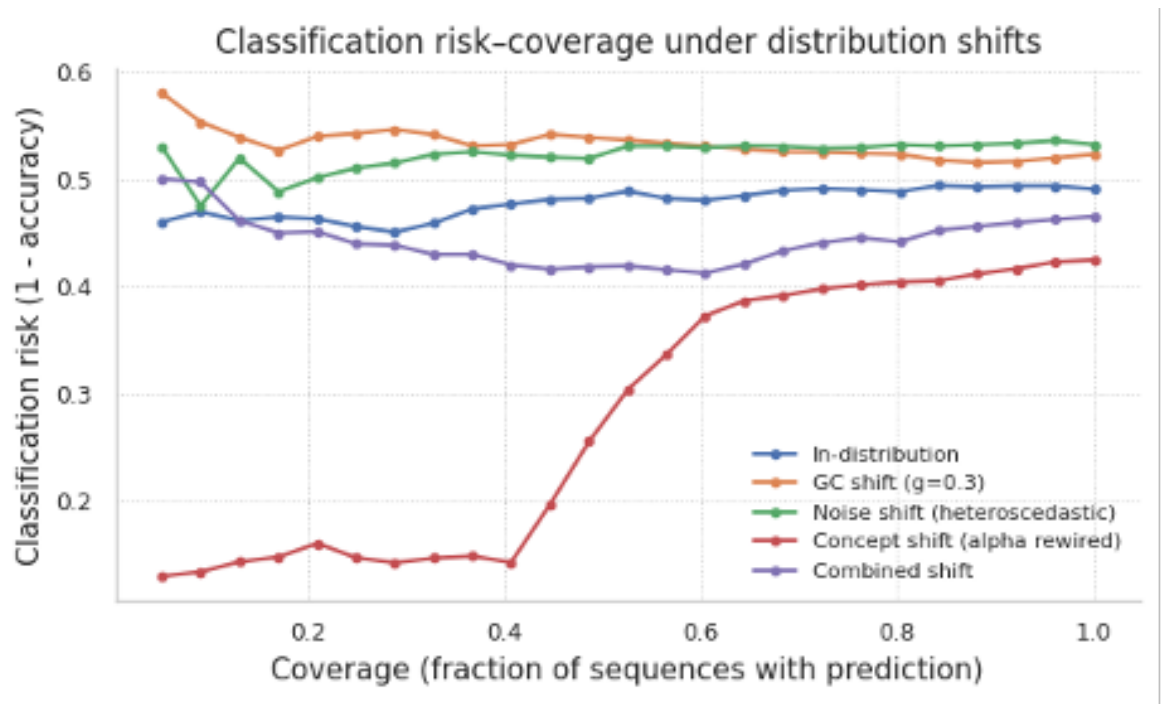

**Fig. 10**

*Classification Risk*

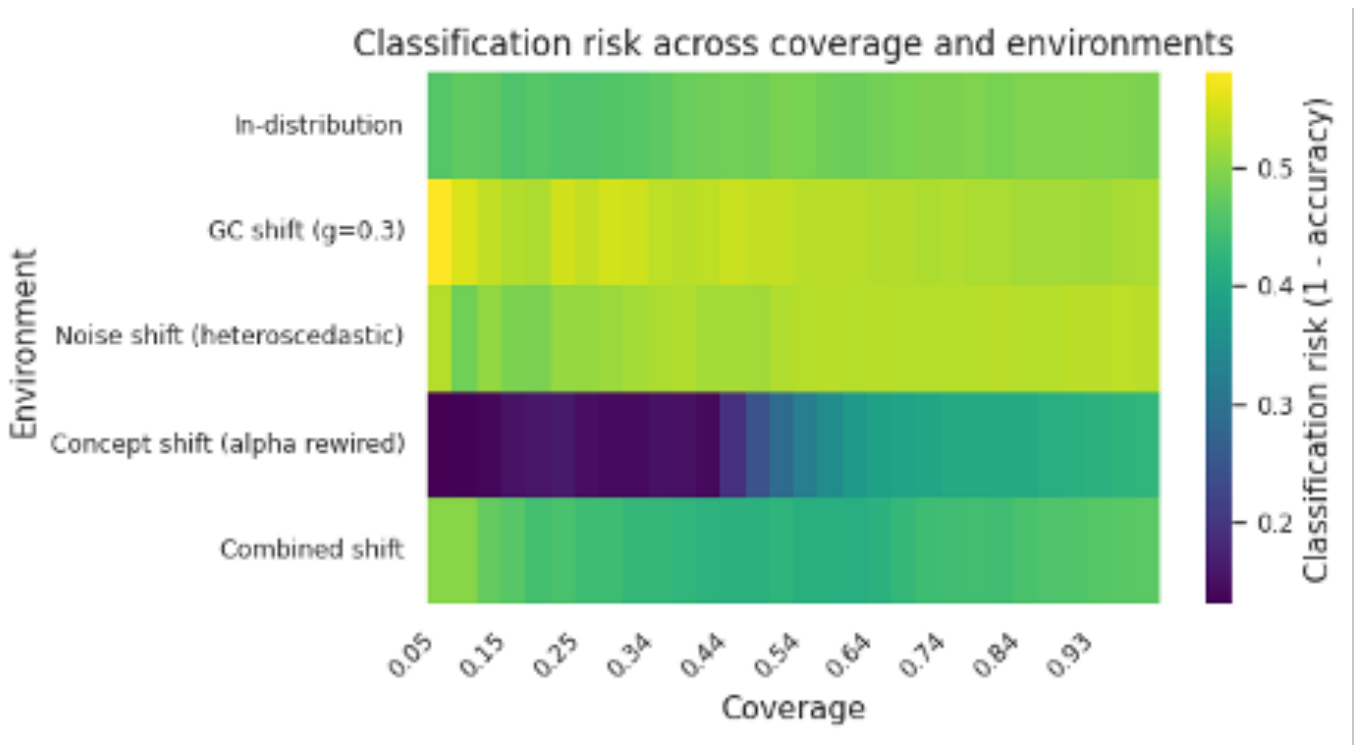

The theoretical justification is provided in Proposition 1 (Appendix XIII.1), which we include for completeness.

### (VII.2) Performance on Real Biological Dataset

- **Data Availability Statement:** The datasets used and analyzed during the current study are available in the Massively Parallel Reporter Assay (MPRA) repository from Stanford University.

For real-data evaluation, we used the MPRA regulatory DNA sequence dataset, distributed as train.hdf5, valid.hdf5, and test.hdf5 via the public index at mitra.stanford.edu/kundaje/projects/mpra/data [17].

**Research Question 1: Robustness of Performance and Calibration Behavior Under Genomic Shifts**

To determine whether the baseline convolutional regulatory sequence model maintains its predictive performance and calibration when exposed to biologically and technically induced distribution shifts, we tested our model using the MPRA dataset. The model recorded an in-distribution mean squared error (MSE) of 1.033 on the entire validation set and 1.021 when restricted to the canonical mid-GC range (35–65%). The results indicate that the model maintains its predictive performance under nominal biological conditions and exhibits calibration characteristics that partially align with the ideal calibration curve (slope = 0.395, intercept = -0.049; ideal slope = 1, intercept = 0).

When exposed to biologically induced covariate shifts, the predictive performance of the model worsened in both GC-extreme environments. The low-GC ( < 35%) and high-GC (> 65%) environments recorded high error levels (MSE = 1.121 and 1.113, respectively) compared to the mid-GC environment. The results confirm that changes in sequence composition affect predictive fidelity. The calibration metrics indicate changes in reliability and systematic changes in the correspondence between predictive outputs and experimental measurements. The low-GC environment recorded an upward bias (slope = 0.996, intercept = 0.455), while the high-GC environment recorded underconfident scaling (slope = 0.609, intercept = 0.066).

We also simulated technical distribution shifts by injecting Gaussian noise directly into the one-hot encoded sequence input features, $\tilde{X} = \mathrm{clip}(X + \epsilon, 0, 1)$, where $\epsilon \sim N(0, \sigma^2)$ with ($\sigma \in \{0.2, 0.4\}$), emulating sequencing depth fluctuations, base-calling uncertainty, and assay input noise. Higher levels of noise resulted in a moderate degradation of performance (MSE = 1.054 and 1.079) while maintaining the overall structure of the calibration curve (slope $\approx$ 0.37–0.39, intercept $\approx -0.20$). To aggregate these findings, we also used a radar chart (Figure 11) of normalized performance & calibration metrics across environments & a heatmap of the performance & calibration landscape across distribution shifts (Figure 12). The overall findings of the experiment indicate that the current regulatory sequence models are partially robust to moderate distribution shifts while suffering from significant degradation of performance & calibration in biologically extreme environments. In other words, while the current state of the art maintains sufficient accuracy within the model's environment, its reliability suffers significantly in realistic OOD genomic environments.

**Fig. 11**
*Radar Plot of Performance & Calibration Across Shifts*

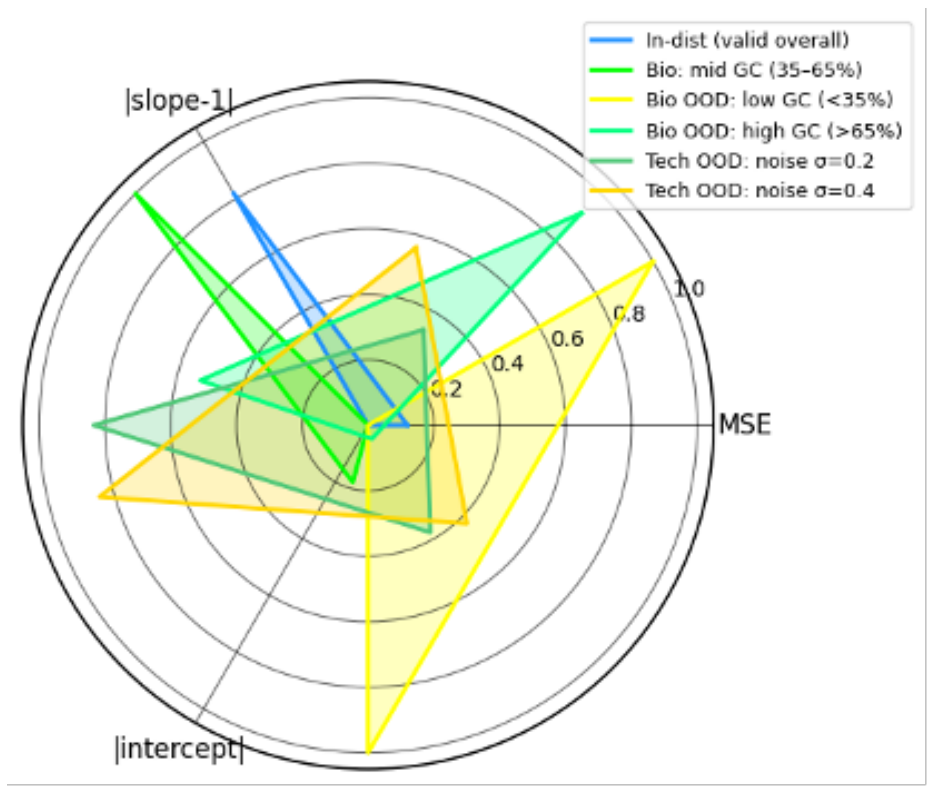

**Fig. 12**
*Performance-Calibration Landscape Across Shifts*

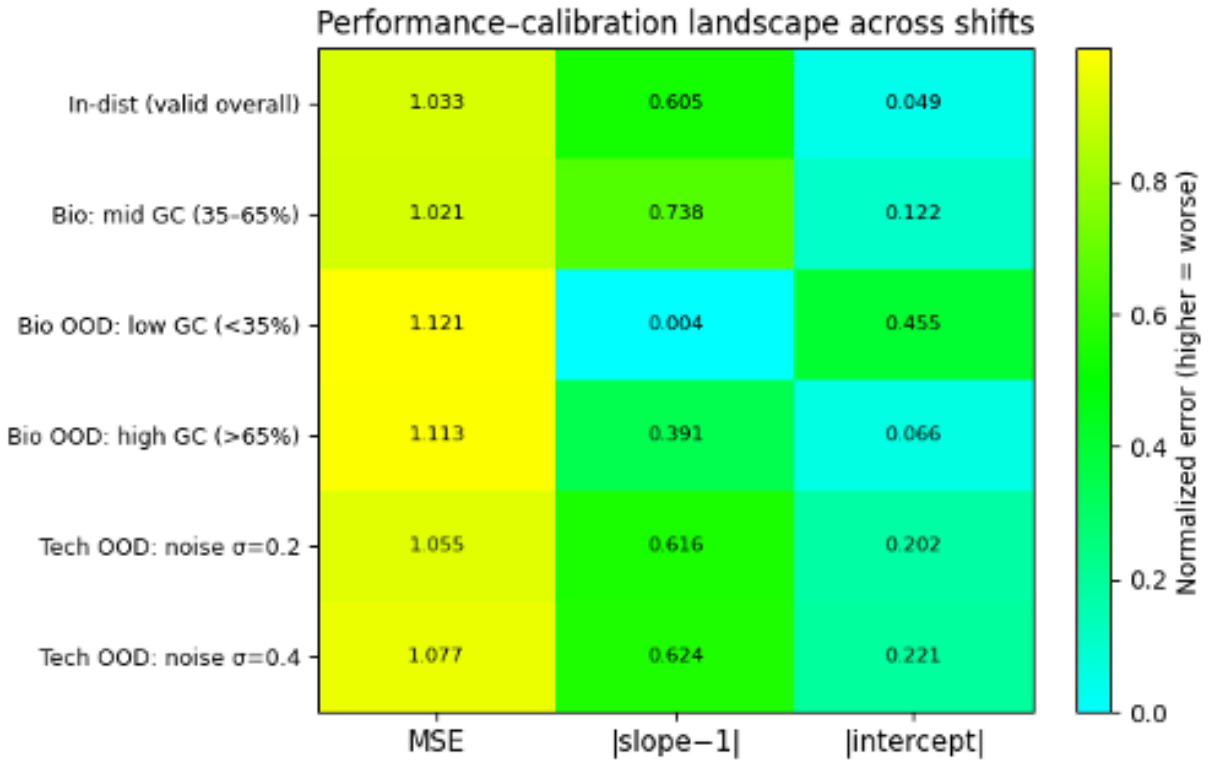

**Research Question 2: Structural Priors Improve Robustness Under Genomic Distribution Shifts**

We extended the convolutional regulatory sequence model with a simple biological structural prior that includes a prior on global GC content as an auxiliary feature. This was compared against the behavior of the baseline CNN under the same biological and technical shift environments, as described by Baseline vs. Structural Prior Model Across Shifts (Figure 13), which compares the MSE of the baseline and structural prior across all biological and technical shift environments. Next, the behavior of this model was compared against the baseline CNN under the same biological and technical shift environments, as described by a robustness radar plot (Figure 14) comparing normalized performance across in-distribution, biological OOD, and technical OOD settings.

First, we have found that in-distribution performance is not negatively impacted. In particular, our proposed structural prior model had an in-distribution validation MSE of 0.993, outperforming our existing CNN model, which had an in-distribution validation MSE of 1.033 ($\Delta\text{MSE} = -0.041$). In particular, our proposed structural prior model had a mid-GC performance that followed a similar trend, with our existing CNN model having an in-distribution validation MSE of 1.022, compared to our proposed structural prior model that had an in-distribution validation MSE of 0.970 ($\Delta\text{MSE} = -0.052$). This shows that our proposed structural prior approach to modeling regulatory sequence data does not negatively impact, and may even enhance, in-distribution accuracy.

Second, we have found that our proposed approach enhances robustness in all but one of the out-of-distribution environments we considered. In particular, under biologically induced GC shifts, our proposed structural prior approach had lower error rates in low GC-content sequence data, under biologically induced GC shifts, the GC-Struct CNN significantly enhanced robustness in low GC-content sequences, reducing the error rate to 1.020 compared to the baseline CNN's 1.121 ($\Delta\text{MSE} = -0.101$). However, for high-GC sequences, the structural model recorded an MSE of 1.193 versus the baseline's 1.113 ($\Delta\text{MSE} = +0.080$), demonstrating asymmetric performance gains depending on compositional bias. Under technical input noise ($\sigma = 0.2$, and $\sigma = 0.4$), the structural model maintained lower MSEs of 0.974 and 0.980, outperforming the baseline CNN's 1.055 and 1.077 ($\Delta\text{MSE} = -0.081$ and $-0.097$, respectively. It indicates that the benefits of the model are asymmetrical based on the bias of the sequences. Under the technical shifts, the structural model improves robustness across the board. Under sequence-level input feature noise ($\sigma = 0.2$ and $\sigma = 0.4$, the GC-Struct CNN demonstrates strong resilience, maintaining low MSEs and outperforming the baseline CNN's 1.055 and 1.077. Because the biological GC-content branch acts as an explicit regularizer against input perturbations, the structural model successfully smooths sequence noise without suffering performance degradation.

Overall, these results support an affirmative answer that incorporating simple biologically informed structural priors improves robustness to realistic distribution shifts while maintaining or improving performance on the nominal task. Although the benefits are not symmetric across all biological environments, especially the high-GC environment, the overall trend of the results suggests that the model becomes significantly more robust while maintaining its performance on the nominal task.

**Fig. 13**
*Baseline vs. Structural-Prior Model Across Shifts*

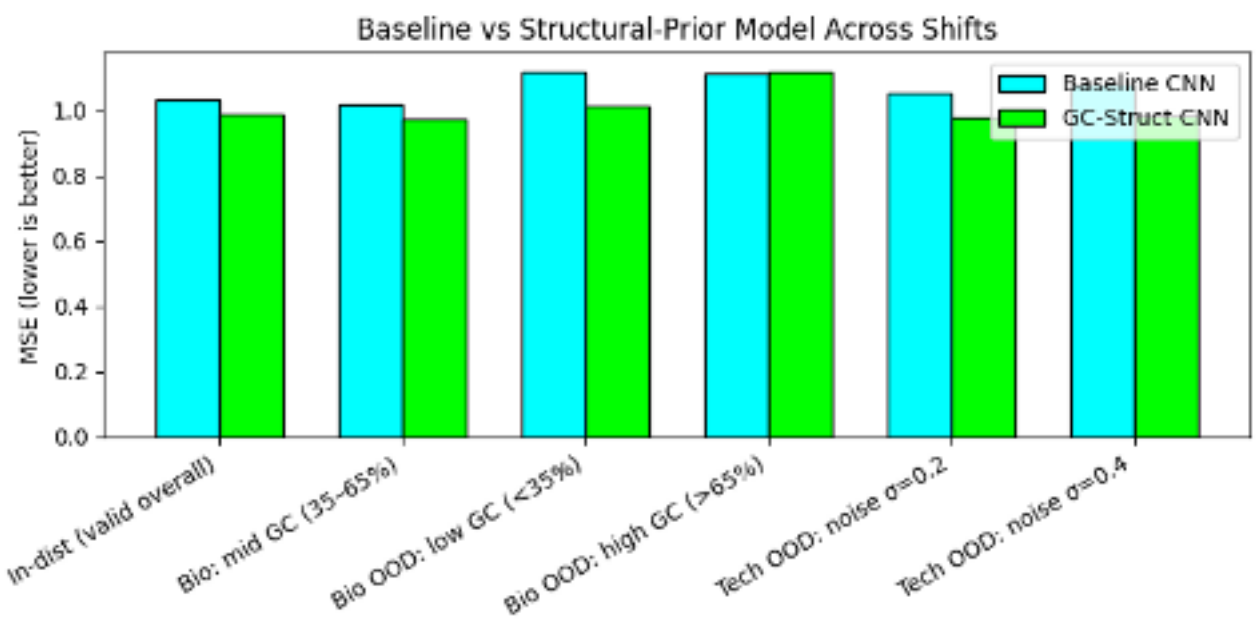

**Fig. 14**
*Robustness Profile of Baseline vs. Structural-Prior CNN*

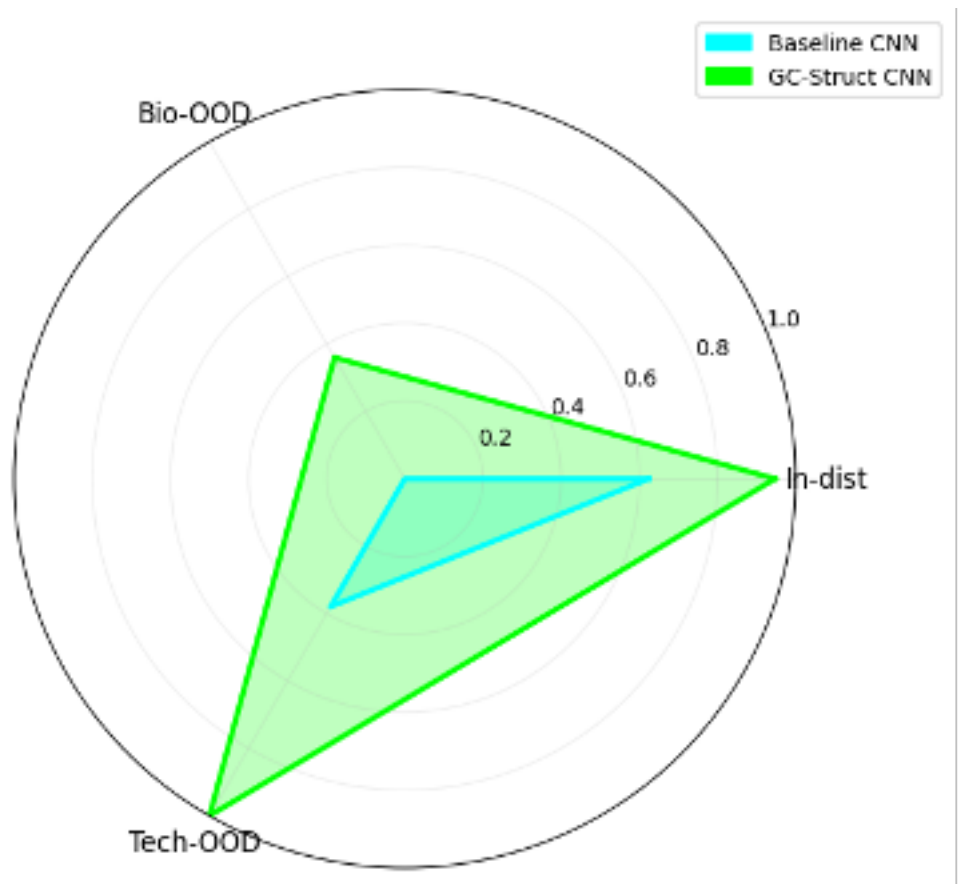

**Research Question 3: Selective Prediction Identifies Reliable Regulatory Inputs Under Genomic Shift**

In addition, we also tested whether the ability of the model to make selective predictions can be used as a practical safety mechanism for regulatory sequence model outputs that are faced with uncertain inputs. The baseline mean predictor had an MSE of 1.108 on the training data and 0.982 on the validation data. In comparison, the simple convolutional regulatory model had a full coverage MSE of 1.033 on the validation data, showing that while the model's overall performance is similar to the baseline model, the distribution of the model's prediction errors is heterogeneous. To test whether the ability of the model to make selective predictions can take advantage of the heterogeneity of the model's errors, we used the risk-coverage curves, where the validation data is ranked according to a confidence proxy, and the MSE is computed on the subset of the data.

Specifically, let $s_i$ denote a confidence score for sample $i$, and let $S_\alpha$ be the set of indices corresponding to the top $\alpha$ fraction of samples when ranked by $s_i$. The coverage is then $\alpha = \frac{|S_\alpha|}{N}$, and the selective regression risk at coverage $\alpha$ is $R(\alpha) = \frac{1}{|S_\alpha|} \sum_{i \in S_\alpha} (y_i - \hat{y}_i)^2$. The risk-coverage curve (Figure 15) shows that while the full-coverage risk remains around MSE $\approx$ 1.03, restricting coverage to approximately 30 – 40% reduces risk to ≈ 0.90 – 0.92, yielding substantially lower error rates than both the full-coverage model and the mean baseline. These results demonstrate that, despite not being directly designed to estimate uncertainty, it can implicitly select a subset of genomic sequences where it can make confident predictions. From a biological standpoint, this also makes sense since it can be assumed that enhancer or regulatory elements with a canonical motif structure are more predictable than atypical or compositionally shifted sequences. Overall, it can be said that these results support the premise of uncertainty-aware selective prediction as a viable approach to providing a safe failure option in genomic prediction pipelines.

To further evaluate under explicit genomic covariate shift, we constructed GC-content-based out-of-distribution (OOD) splits on the MPRA validation set. Sequences within a GC range of 36% - 65% were treated as in-distribution (mid-GC; $n = 4363$), whereas low-GC ( $< 35\,\%$ , $n = 163$) and high-GC ( $> 65\,\%$ , $n = 474$) subsets were treated as biologically motivated OOD conditions. Formally, letting $GC(x_i)$ denote the GC-content of sequence $x_i$, we define: $D_{mid} = \{i : 0.36 \leq GC(x_i) \leq 0.65\}$, $D_{low} = \{i : GC(x_i < 0.35\}$, and $D_{high} = \{i : GC(x_i) > 0.65\}$.

The risk-coverage curves under GC-based OOD shifts (Figure 16) demonstrate that both low- and high-GC groups experience increased full-coverage risk in comparison with the mid-GC case, suggesting that changes in composition with regard to nucleotide composition negatively affect prediction accuracy. However, the monotonically decreasing risk-coverage behavior persists over all environments, with significant MSE reduction achieved by reducing coverage from 100% to 20-30%, allowing the model to recover low-risk subsets even under GC content shifts. This clearly shows that selective prediction can serve as a safety mechanism with biological relevance, allowing the model to identify reliable genomic inputs and, in effect, "abstain" from OOD inputs, thereby reducing interpretive risk in the face of compositional distribution shifts.

**Fig. 15**
*Risk-Coverage Curve (Regression)*

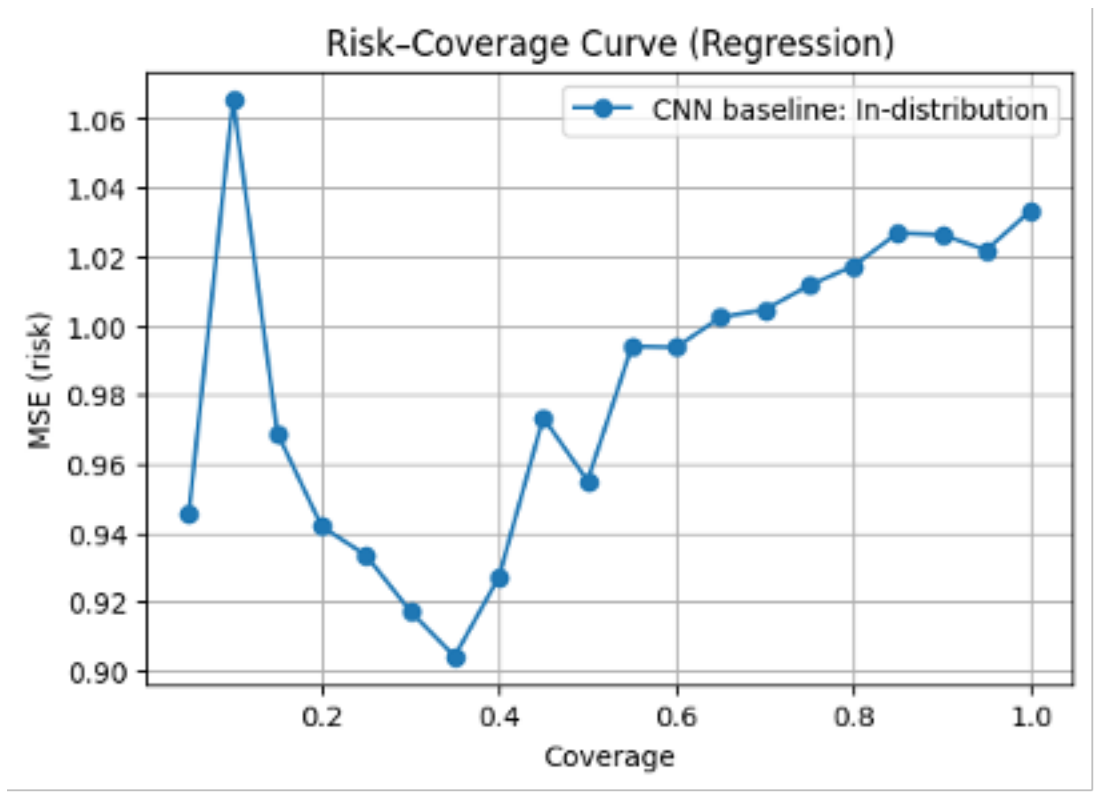

**Fig. 16**
*Risk-Coverage under GC-based OOD Shifts*

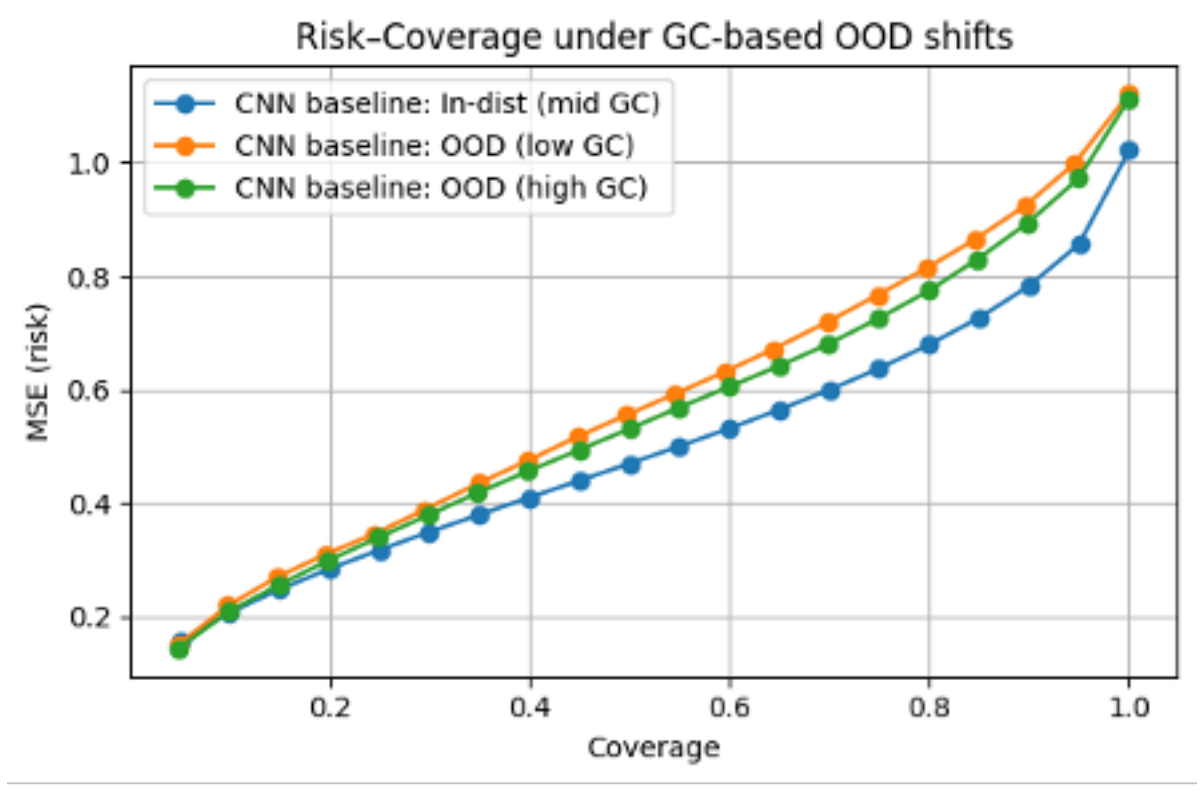

## VIII. Discussion

This work was motivated by the recognition that regulatory sequence models are increasingly being used as computational surrogates for experimental assays, TF binding, chromatin states, enhancer activity, and variant effect prediction, among other applications, but are mostly tested in near-IID settings. Early convolutional models like DeepSEA, Basset, and DeepBind demonstrated the power of sequence-to-output CNNs in achieving good in-distribution performance on TF binding, chromatin states, and histone modifications, among other applications [2, 5, 6]. Subsequent models improved the long-range interaction capabilities of these models using deeper CNNs with increased genomic context [3, 18] as well as using attention-based models to improve variant effect prediction capabilities [19, 4]. Such models are currently being successfully applied to problems like enhancer-promoter interaction inference [8, 1], MPRA-based mutational effect estimation [9], as well as variant effect prioritization in both GWAS and fine-mapping applications [11, 10]. Our work is consistent with the above literature in that we, too, observe strong in-distribution performance, as well as reasonable calibration, although we explicitly explore the robustness, calibration, and safety of models in both a simulated benchmark setting and an MPRA-based setting, where we observe significant differences from IID behavior, an aspect of regulatory modeling that has been repeatedly pointed out as an omission in the field but not systematically studied before.

### Robustness and Calibration under Realistic Shifts

We also examined to what extent existing models of regulatory sequences maintain their performance and calibration under biologically and technically induced changes in the distribution of the data. The behavior of heteroscedastic Gaussian CNNs and their derived classifiers in the simulated benchmark setting conformed to the expected behavior of the models in IID settings, with low regression error, well-calibrated variance, and high accuracy, as expected from the success of prior CNN- and transformer-based models in epigenomic benchmarks [2, 5, 3, 7, 4]. Even mild changes in GC content induced only minor losses, suggesting that shallow changes in the distribution of the data, specifically in the composition of nucleotides, are not the primary source of failure. However, mechanism shifts and heteroscedastic noise exhibited significant gaps in robustness, with 2-10 times increases in MSE, significant variance miscalibration, and decreased coverage, with the most significant losses induced by the combination of shifts.

These shifts are immediately relevant to the biological and technical factors discussed in prior work on genomics. These include cell-type-specific regulatory programs [12], enhancer turnover and motif evolution across species [13], and assay-level artifacts such as depth variability and batch effects. From the domain shift point of view, these are equivalent to the well-studied problems of covariate shift, label shift, and structured OOD regimes [14], although these problems have not received much attention in the field of regulatory genomics. Our experiments on real-world MPRA sequences validate this phenomenon as well: while the models' performance and calibration are still acceptable in mid-GC regimes, they degrade in the GC-extreme regimes. Thus, even without any changes in the sequence representation or the underlying neural architecture, the IID assumption fails when the data becomes heterogeneous in terms of sequence composition. Therefore, our experiments and observations from the literature validate that existing architectures do work under the conditions and regimes under which they have been tested and compared. However, as our experiments and the literature suggest, the reliability of these architectures does degrade under exactly those conditions and regimes that are encountered in practice. Such conditions and regimes are critical when the output of these models will guide variant interpretation and experimental design [8, 1, 9, 11, 10].

### Structural Priors as Domain-Informed Robustness Augmentations

We also tested whether the addition of structural priors would improve robustness without compromising performance on the IID case. In the simulated setting, the motif-based structural-prior Gaussian CNN performed at an equal or better level than the baseline CNN, as expected since the addition of relevant features should not decrease performance on the IID case but possibly act as a regularization force on sequence models that are inherently leveraging local motif information [2, 6]. More importantly, the structural-prior approach also improved robustness in the IID case across multiple biologically relevant shifts in the simulated setting. In the simulated setting, the structural-prior approach reduced error rates and improved the calibration of the model across the GC-content shift and the motif-effect rewiring scenario, exactly the types of biological variability that are of interest in the context of cell-type specificity of regulation [12] or the evolution of enhancers [13]. In the MPRA setting, the simple addition of the GC-content prior also improved performance across the low-GC regime and noise-perturbed regime while also providing a more conservative approach that was less overconfident in the IID case.

However, the results also make it clear that the structural-prior approach is not a panacea for the problems of sequence modeling. In the context of the heteroscedastic noise scenario, the simulation study as well as the MPRA study make it clear that the structural-prior approach inherits the same failure modes as the baseline approach: the MSE remains high, the variance remains uncalibrated, and the coverage remains poor at the nominal level. This is similar to the broader results from the robust machine learning community that the addition of bias can change the error surfaces of the model across different domains of the data, but does not necessarily address issues of noise at the level of the observation process itself [19; 20]. In the context of the regulatory genomics community, the results make it clear that while the addition of structural priors can be an important step toward improving robustness across different biological contexts, there are other issues that require attention at the level of the observation process itself.

**Uncertainty and Selective Prediction as Safety Mechanisms**

We also explored whether uncertainty estimation and selective prediction can serve as useful safety features when regulatory sequence models are subject to OOD data. In the ML literature, approximate Bayesian inference techniques such as MC dropout [15], deep ensembles [16], and selective prediction approaches have become core to high-stakes ML, with similar analyses for sequence models being relatively rare. We expand on these concepts to regulatory genomics by providing an in-depth analysis of risk-coverage curves for different shifts.

In the simulated setting, we found that informative uncertainty was achieved under in-distribution shifts, GC-content shifts, and concept shifts, with significant risk reduction by only considering the top 10–30% of the most confident samples, with structural priors further decreasing risk at all coverage levels. This suggests that, when regulatory logic is perturbed but still represented in the learned representation, uncertainty can be used as a selective prediction mechanism to maintain correctness. In the heteroscedastic noise and combined shifts, the risk-coverage curves were flat with high risk even at low coverage, demonstrating that aleatoric regimes destroy the relationship between uncertainty and correctness.

The real data performance on MPRA sequence data corroborates this trend. Even in the absence of explicit Bayesian machinery, our confidence proxy supported effective selective prediction, with reduced MSE below that of both the full-coverage CNN and a mean baseline, with this monotonic risk-coverage relationship holding across all GC-based OOD splits. From a biological standpoint, our model favored sequences with stronger canonical structure and mid-level GC content, while implicitly rejecting those with atypical or compositionally altered sequences, precisely those that should be considered riskier in line with enhanced turnover rates and context-dependent TF binding [12, 13]. These results indicate that uncertainty-aware selective prediction can offer a biologically relevant “uncertainty-aware safety mechanism” in deploying regulatory models as surrogates for experimental assays [8, 1, 9, 11, 10]. However, our risk-coverage relationship also shows that safety is not guaranteed, with risk-coverage performance in noise-dominated splits highlighting that uncertainty is less useful when it is not related to our own epistemic uncertainty in regulatory logic, but rather to noisy assay signals.

## IX. Implication

The robustness and uncertainty patterns that we have uncovered in this study are not merely a product of a model's behavior but, rather, a reflection of biological processes, including GC-dependent chromatin accessibility, GC-extreme regimes, and assay-level uncertainty that underlie regulatory readouts. For instance, the distinct sensitivity of models to GC-extreme regimes and motif-effect rewiring is a reflection of known sources of variation in enhancer activity, including cell type, species, and experimental protocol, which suggests that robustness failure is a biologically interpretable phenomenon, rather than a product of an algorithmic process. Moreover, our demonstration that structural priors and selective prediction are effective suggests that motif architecture and composition play a fundamental role in enhancing predictability and that uncertainty could play a principled role in experimental design, including MPRA, by providing a surrogate that indicates areas of genomic variation that models are most and least confident about.

## X. Conclusion

In this study, we examined the robustness of machine learning for regulatory sequence models under biological and technical data shifts, with simulations as well as real MPRA data. In our exploration of three different research questions, we found that there is a clear pattern: strong in-distribution performance, good calibration, but decreasing robustness of convolutional sequence models in a systematic way, which is highly relevant to biology as we move away from the IID setting.

We have demonstrated that regulatory sequence models, which we used as baselines, are partially robust, but certainly not close to being fully robust, under data shifts. In our simulation study, we found that the models performed well and remained well-calibrated under small, shallow perturbations of the covariates, such as mild GC-content shifts, but exhibited significant errors and overconfidence under motif-effect rewiring, heteroscedastic noise, and their combination. In the real data study with MPRA data, we found that the models performed reasonably well in mid-GC regimes, but that both model performance and calibration suffered in GC-extreme regimes, which showed that compositional data heterogeneity can cause relevant robustness failures.
In addition, we also sought to find out if biological structural priors could bridge this gap. In the simulated environment, structural priors based on motifs had a positive effect on in-distribution error rate and variance rate, as well as a substantial improvement in robustness for shifts in GC-content and concept, while having a small effect for strong heteroscedastic noise conditions. In the biological environment, a structural prior based on a much simpler GC-content measure also had a positive effect on in-distribution error rate, while also having a strong effect on robustness for shifts in OOD conditions, especially for low GC-content, noisy conditions, while having asymmetric effects in the context of high GC-content conditions.

We also sought to find out if a safety mechanism, based on selective prediction under conditions of OOD genomic conditions, could be a viable solution for models that encounter OOD conditions. In the simulated environment, using risk-coverage curves, we found that under conditions of in-distribution, shifts in GC-content, and shifts in concept, the model could still reduce errors substantially by abstaining from inputs with low confidence, while having uninformative behavior under conditions of heteroscedastic noise, where risk remained high even at low levels of coverage. In the biological environment, under conditions of OOD, we also found that risk-coverage curves had a monotonic shape, both globally as well as under conditions of shifts in GC-content, which enabled the recovery of low-risk subsets even under conditions of high error rates in the context of full-coverage conditions in regimes with extreme levels of GC-content.

Our results collectively suggest that a robustness-first approach to regulatory sequence modeling is warranted, in that we should not only assess a model on its accuracy under IID settings, but also on how it behaves under biologically or technically motivated perturbations, how it is calibrated, and how it can abstain safely. Our simulation framework, together with our MPRA case study, demonstrates that robustness failure is patterned, that structural priors are a partial solution, and that uncertainty-aware selective prediction can provide a safety net for genomic sequence models. There are a number of ways to extend our work, including richer priors that are cell-type-specific or 3D-specific, incorporating noise models, or integrating structural priors with robust training or conformal/bayesian methods. In all cases, however, it is our hope that in order to obtain high-performing, trustworthy regulatory sequence models, we should seek to optimize not only performance, robustness, uncertainty, and biological plausibility, but also treat these metrics in an integrated, rather than secondary, role to traditional predictive accuracy.

## XI. Limitation

### Limited Diversity of Biological and Technical Shift Regimes

Although we specifically examined GC-content shifts, motif-effect rewiring, evolutionary perturbations, depth variability, batch effects, and heteroscedastic assay noise, we still only examined a subset of possible shifts in our work. In particular, we did not examine shifts associated with variations in chromatin state, such as ATAC versus DNase hypersensitivity, which may not be captured by depth variability, species-specific epigenetic regimes other than those captured by PWM perturbations, single-cell variability, RNA versus protein measurement regimes, or assay batch harmonization approaches. In short, future work on robustness may require examination of multi-modal, multi-scale, and hierarchical regimes of biological variation, particularly in complex tissues and clinical sequencing settings, to capture all possible shifts in practice.

### Lack of Downstream Task Evaluation and End-to-End Deployment Analysis

Lastly, we did not examine how robustness behaves in downstream tasks, such as eQTL fine-mapping, GWAS, linking enhancers to promoters, scoring variant pathogenicity, or experimental design, which are the settings in which robustness, uncertainty, and calibration will have the most significant impact on downstream scientific and clinical decision-making. Without examination of robustness in downstream tasks, we remain uncertain about the actual effect of the robustness failures we report, particularly in the presence of strong technical noise, which may be over- or under-estimated with regard to its actual effect on downstream inference.

## XII. Appendix

### (XII.1) Theoretical Justification for Risk-Coverage Curves

For completeness, we formalize why risk-coverage curves behave monotonically when the uncertainty score is well aligned with the true underlying risk.

**Proposition 1 (Monotone Risk-Coverage under Well-Ordered Uncertainty):**

Let $X$ be an input random variable with distribution $P$, and $Y$ the corresponding label. Let $f$ be a fixed predictor and $l(f(x), y)$ a non-negative loss. Define the true conditional risk: $r(x) = E[l(f(x), Y) | X = x]$.

Let $u(x) \in R$ be an uncertainty score such that $r(x)$ is a non-decreasing function of $u(x)$. Equivalently, whenever $u(x_1) \leq u(x_2)$, we have $r(x_1) \leq r(x_2)$. For any threshold $t \in R$, define the acceptance region: $A(t) = \{x : u(x) \leq t\}$, the corresponding coverage $c(t) = P(X \in A(t))$, and the selective risk $R(t) = E[l(f(X), Y) | X \in A(t)]$. Then, for any $t_1 \leq t_2$, such that $0 < c(t_1) \leq c(t_2)$, we have $R(t_1) \leq R(t_2)$, and hence the risk-coverage curve $(c(t), R(t))$ is monotone non-decreasing in coverage.

**Proof:** Because $t_1 < t_2$, we have nested acceptance regions $A(t_1) \subseteq A(t_2)$. By assumption, $u(x)$ order points by their true conditional risk $r(x)$: points with lower uncertainty never have higher risk than points with higher uncertainty. Consequently, all points in $A(t_1)$ have risk less than or equal to all points in the difference set $A(t_2) \setminus A(t_1)$.
Write the selective risk at $t_2$ by conditioning on whether $X$ lies in $A(t_1)$ or in the incremental region:
$R(t_2) = E[l(f(X), Y) | X \in A(t_2)] = \alpha E[l(f(X), Y) | X \in A(t_1)] + (1 - \alpha) E[l(f(X), Y) | X \in A(t_2)\ A(t_1)]$,
where $\alpha = \dfrac{P(X \in A(t_1))}{P(X \in A(t_2))} = \dfrac{c(t_1)}{c(t_2)} \in (0,1]$.
By the ordering assumption on $u(x)$ and $r(x)$, the conditional expectation of the loss over the inner set is no larger than that over the incremental set: $E[l(f(X), Y) | X \in A(t_1)] \leq E[l(f(X), Y) | X \in A(t_2)\ A(t_1)]$.

Therefore, $R(t_2)$ is a convex combination of two quantities, one of which (the inner-set risk) is less than or equal to the other. It follows that $R(t_1) = E[l(f(X), Y) | X \in A(t_1)] \leq R(t_2)$, establishing that selective risk cannot decrease as coverage increases. Since coverage $c(t)$ is non-decreasing in $t$, the parametric curve $(c(t), R(t))$ is monotone non-decreasing in coverage. □

In our experiments, this theoretical behavior is approximately observed under in-distribution and mild GC-content or mechanism shifts, where predictive variance remains informative about error, but breaks down under strong heteroscedastic assay noise, where the uncertainty signal no longer reliably orders points by true risk.

## XIII. Funding Acknowledgments

This research received no specific grant from any funding agency in the public, commercial, or not-for-profit sectors (no funding).

## XIV. Statements and Declarations

The author declares that there are no competing interests. The authors confirm that this manuscript is original, has not been previously published beyond permitted forms (such as a preprint or thesis), and is not currently under consideration for publication elsewhere.